\documentclass{PoS}

\title{Preliminary results of giant pulse investigations from Crab pulsar with Radioastron.}

\ShortTitle{Preliminary results of giant pulse investigations from Crab pulsar with Radioastron.}

\author{\speaker{Rudnitskiy A. G.}\\
        Astro Space Center, Lebedev Physical Inst. RAS, Profsoyuznaya 84/32, 117997 Moscow, Russia\\
        E-mail: \email{almax1024@gmail.com}}
\author{Popov M. V.\\
        Astro Space Center, Lebedev Physical Inst. RAS, Profsoyuznaya 84/32, 117997 Moscow, Russia\\
        E-mail: \email{mwpopov@gmail.com}}
\author{Soglasnov V. A.\\
        Astro Space Center, Lebedev Physical Inst. RAS, Profsoyuznaya 84/32, 117997 Moscow, Russia\\
        E-mail: \email{vsoglasn@asc.rssi.ru}}

\abstract{Giant pulses from Crab pulsar were observed together with Radioastron space radiotelescope, Global EVN radio telescopes and Kvazar-KVO VLBI stations to study the scattering effects in the ISM. Five observing sessions were conducted at 18 cm (EVN codes: EG060A, EG060B, EG067B, EG075) and one at 92 cm (EVN code: GS033A). We have estimated distance to the scattering screen, angular size of scattering disk, scattering time. All estimations were done in an assumption of single thin scattering screen. Also, a significant change in the shape of cross-correlation functions for space-ground baselines was found (starting from 4 up to 12 Earth diameters).}

\FullConference{12th European VLBI Network Symposium and Users Meeting\\
                 7-10 October 2014\\
                 Cagliari, Italy}

\begin{document}

\section{Introduction.}
Pulsar scintillation is an excellent probe of intrinsic pulsar physics and of the turbulent interstellar plasma. The scintillation can be studied from a variety of perspectives, including its autocorrelation (in frequency, time, or space), the temporal broadening, and the angular broadening. Each of these provides a critical element for developing a comprehensive understanding of the scattering. For instance, the temporal broadening can indicate whether the scattering material is localized along the line-of-sight but cannot determine the position of the scattering material without measurements of angular broadening. 

The spatial autocorrelation function, measured by an interferometer, can provide information that is inaccessible by other means. Even
in the most basic case, the intensity temporal autocorrelation function is only sensitive to the squared spatial autocorrelation, enabling interferometry to more delicately reflect small-scale structure. In addition, interferometry is crucial when the scattering or source properties change over short timescales. Hence, space-VLBI provides an irreplaceable tool for studying giant pulses from the Crab pulsar and their complex associated scattering environment.

The Crab pulsar is unusual among pulsars in that most of its radio emission arises from giant pulses: narrow, strong pulses with $10^2$ to $10^7$ times the flux density of an average pulse \cite{S1}. Giant pulses are much shorter than a typical pulsar pulse. They have duration of up to 100 $\mu sec$, but consist of many narrow bursts \cite{S2}. The fractional bandwidths of giant pulses are $~0.5$ \cite{S3}, and their distribution of energies follows a power-law \cite{S4, S5}. Giant pulses become more common, although less remarkable, at longer wavelengths. Popov suggests that emission of the Crab consists entirely of giant pulses, drawn from a distribution that varies across the pulse \cite{S4}. Lyutikov and Parikh \cite{S6} and Jessner et al. \cite{S5} suggest that some properties of giant pulses arise from propagation within the pulsar's magnetosphere, or differences in spatial coherence at the point of emission. 

Because of scattering, a single sharp pulse emitted at the pulsar will arrive over a scattering time $\tau$ at the Earth. Because of phase differences, paths with the same travel time cancel or reinforce so that the original sharp pulse gains a complicated, spiky character: "an impulse-response function" \cite{S7}. More generally, propagation convolves any time-series of emission with this function. The Fourier transform of the impulse-response function is the well-known scintillation spectrum.

At a single frequency, the phase and amplitude of the response change with position, as relative length of paths change, producing a diffraction pattern at the observer with a characteristic spatial scale $S_{ISS}$. Multiple points of emission at the source produce superposed diffraction patterns at the observer: scattering acts as a corrupt lens, to produce a corrupt image in the observer plane.
\\
\\
\section{Observations.}
The high flux density of the Crab pulsar's giant pulses and the compactness of their source makes them attractive objects for study on the longest baselines. Indeed, the strongest such pulses are visible by the Radioastron spacecraft in a single-dish mode. A program of observations of these pulses, starting in the fringe-search phase of the mission, has led to observations of a handful of giant pulses, with surprising and sometimes puzzling properties. Important peculiarity of giant pulses interferometry is that one measures an instant visibility function with no averaging. Such circumstances provide particular mode of observing visibility scintillations interesting for theoretical intepretation of scattering. Goodman and Narayan theory of scatter-broadened images gave us at least two interesting ranges of baselines for Crab giant pulse interferometry, the first is the range of baselines where the transition of cross correlation function (CCF) shape occurs, the second is the range of larger baselines where stable visibility can be observed, but with CCF seriously affected by scattering.\cite{S8} In this case, each, even a strong giant pulse, is very important as a statistical data for the distribution of visibility from baseline projection.

Five observations of Crab pulsar were conducted during the early science program and key science program (AO-1) of Radioastron mission. Four other observations were done at 18 cm (EVN codes: EG060A, EG060B, EG067B, EG075) and one at 92 cm (EVN code: GS033A). Observations were conducted together with Radioastron, Global EVN radio telescopes (Ef, Wb, Mc, Ar, Jb, On, Hh, Ur, Nt, Ro, Tr) and Kvazar-KVO VLBI stations. Correlation for space-ground baselines was found in four sessions at 18 cm. Unfotrunately, at 92 cm no correlation with Radioastron was detected.

\section{Data processing and reduction.}
All observations were processed with the Astro Space Center correlator \cite{ASC1}. An additional mode was developed for giant pulses. This mode is based on the regular pulsar correlation implemented in ASC correlator, where incoherent dedispersion is applied. To search and to correlate giant pulses, after the incoherent dedispersion is applied for a given data window, correlator performs inverse FFT to obtain the cross-correlation function for each baseline. A criteria for each baseline, i.e. whether fringe amplitude is above the given treshold, is then being checked. In case the fringe was detected the correlator writes down the data for all baselines for a given data window. Such approach provides detection for giant pulses that have correlation exceeding the treshold of $6.0$ SNR.

Due to the fact that correlated amplitude for each giant pulse is diffrenet compared to other detected pulses, amplitude normalization is required to obtain correct visibility. The following method was used to perform amplitude correction. Generally, $A_{Norm}= A / R$, where $R$ is a coefficient that can be calculated in two ways.

For space-ground baselines:
\begin{eqnarray}R = (\sigma^2_{2tot}-\sigma^2_{2off}) \cdot \frac{\sigma_{1off}}{\sigma_{2off}}\sqrt{\frac{SEFD_2}{SEFD_1}},\end{eqnarray}

where "1" -- space radiotelescope, "2" -- ground telescope.

For ground-ground baselines:
\begin{eqnarray}R = \sqrt{(\sigma^2_{1tot}-\sigma^2_{1off})\cdot(\sigma^2_{2tot}-\sigma^2_{2off})},\end{eqnarray}

where "1" and "2" -- ground telescopes; $\sigma_{tot}=\sigma_{on}-\sigma_{off}$, $\sigma_{off}$ -- average noise level off pulse, $\sigma_{on}$ -- average level of signal on pulse.

\section{Preliminary results}
We have estimated several parameters to get primary information about scattering in each observation: scattering time, angular size of scattering disk, distance to the scattering screen and diffraction radius. Our preliminary estimations were focused on obtaining information about the distance to the scattering screen. A single thin scatering screen model was used. 

\begin{figure}[h]
\begin{center}
\includegraphics[width=0.3\linewidth]{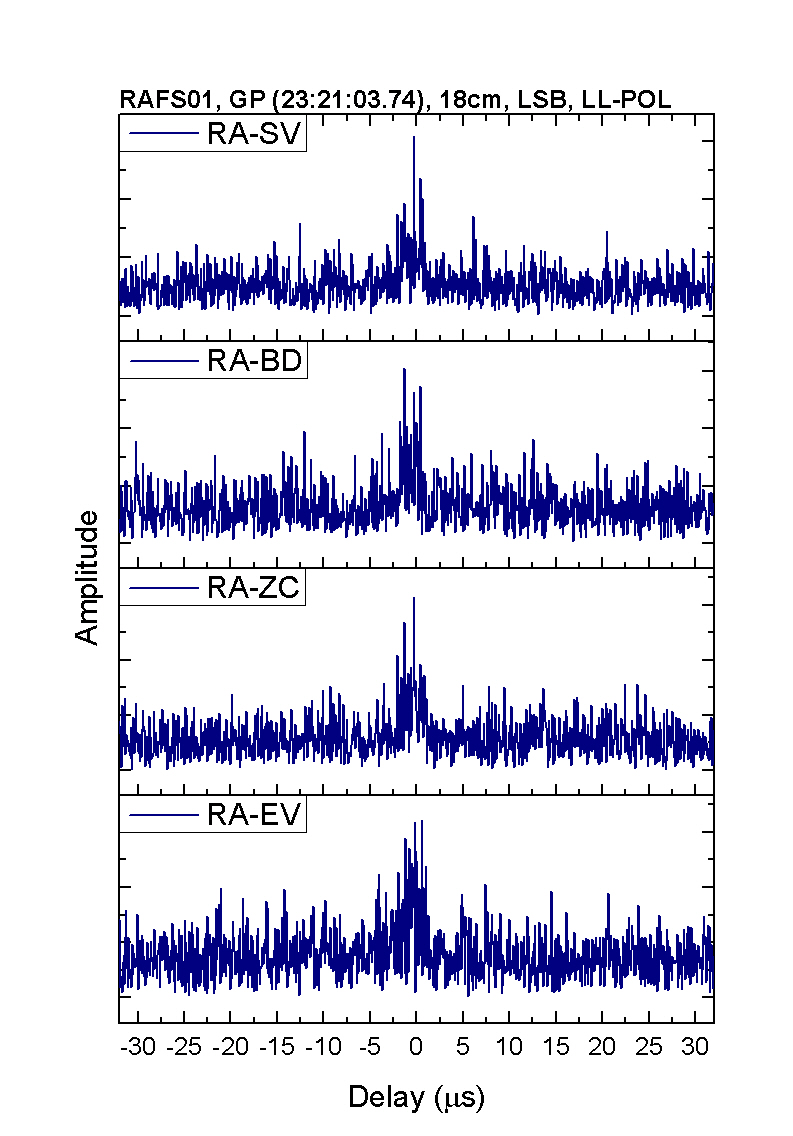}
\includegraphics[width=0.3\linewidth]{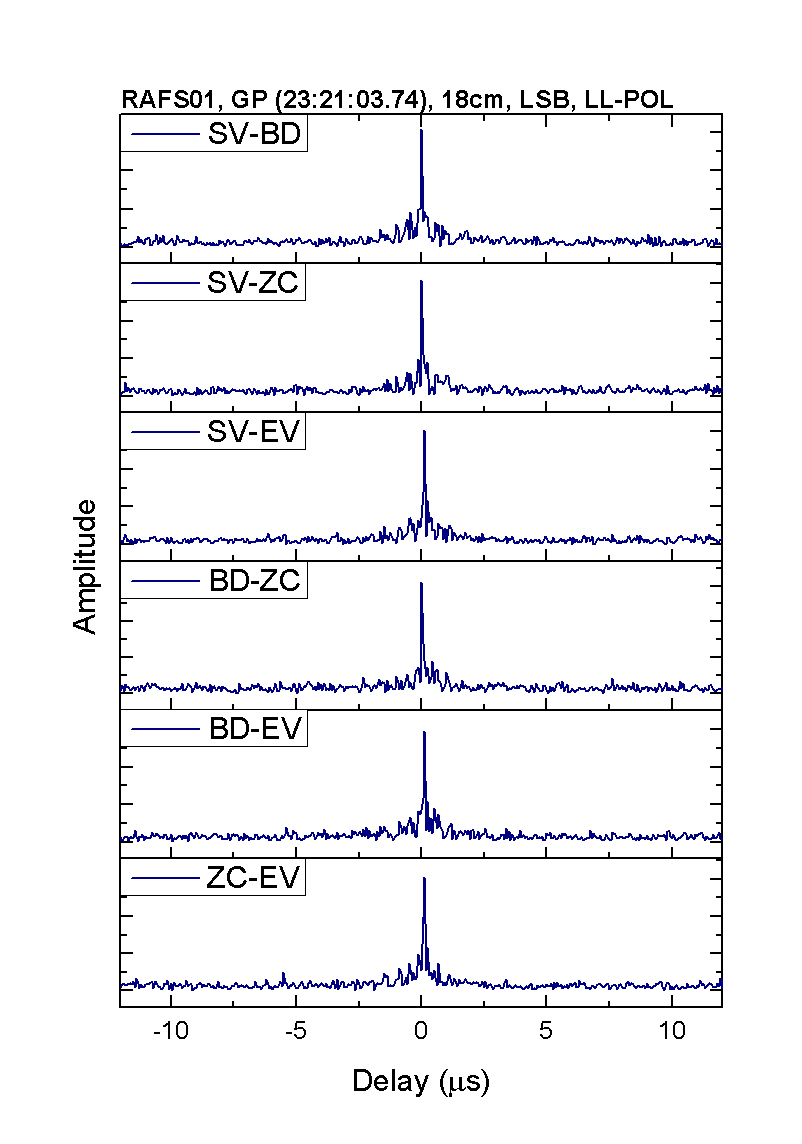} 
\caption{Shapes of visibility amplitude for space-ground baselines (left) and ground-ground baselines (right) for observation RAFS01 (14.11.2011).}
\end{center}
\end{figure}

The first result that was noticed was the significant change in the visibility shape going from the ground-ground baselines up to the larger baselines -- space-ground (see Fig. 1).

For a given angular broadening visibility amplitude depends on baseline projection according to the following expression. \cite{G1}
\begin{eqnarray}V(B)=V_{0}\cdot e^{-\frac{1}{2} \cdot \big( \frac{\pi}{\sqrt{2 \ln{2}}} \cdot \frac{\theta \cdot B}{\lambda} \big)^2},\end{eqnarray}
where $\theta$ -- FWHM angular diameter, $B$ -- baseline projection in number of wavelengths, $\lambda$ -- observation wavelength.
We used mean amplitude of cross-correlation function for space-ground baselines to estimate the scattering time. We fitted the mean CCF amplitude for space-ground baselines using the expression below:
\begin{eqnarray}A=C+A_{0}\cdot e^{-b\cdot t},\end{eqnarray}
and estimated scattering time $\tau_{s} = 1 / b$, corresponding to CCF level of $1/e$.

With known scattering time $\tau$ and angular size of scattering disk $\theta$ we can estimate the distance from the observer to the scattering screen $\alpha \cdot L$ with respect to the total distance to the pulsar $L$. \cite{G2}

\begin{eqnarray}
\theta^2 = \frac{8c\tau \cdot \ln{2} \cdot (1-\alpha)}{\alpha L}
\end{eqnarray}
and from this relation $\alpha$ is:
\begin{eqnarray}
\alpha = \frac{8c\tau \cdot \ln{2}}{\theta^2 L + 8c\tau \cdot \ln{2}}
\end{eqnarray}

\begin {table}[h]
\caption {Estimated parameters}
\begin {center}
\begin {tabular}{c | c | c | c }
\hline
\hline
Observation date & Scattering Disk Size, (mas) & Scattering Time, ($\mu s$) & Distance, ($\alpha$) \\
\hline
14.11.2011 & 1.318 & 0.9 & 0.36\\
02.03.2012 & 0.618 & 5.8 & 0.94\\
06.03.2012 & 0.501 & 5.5 & 0.96\\
23.10.2012 & 1.182 & 5.1 & 0.79\\
02.11.2013 & 1.235 & 2.2 & 0.61\\
\hline
\end {tabular}
\end {center}
\end {table}

The list of obtained parameters for each observation is shown in the Table 1. 
The value of $\alpha$ changes and is differenet in the observations, at $\alpha>0.95$ effective scattering screen seems to be located close to the Crab nebula. This suggests the need to use multiple screen model to describe both scattering effects located within the Crab nebula and in the extended ISM. For the observation of 14.11.2011 we got $\alpha=0.36$ which means that the scattering situation was equivalent to an homogeneous distribution of scattering material along the line of sight. \cite{G3}

\begin{figure}[h]
\begin{center}
\includegraphics[width=0.49\linewidth]{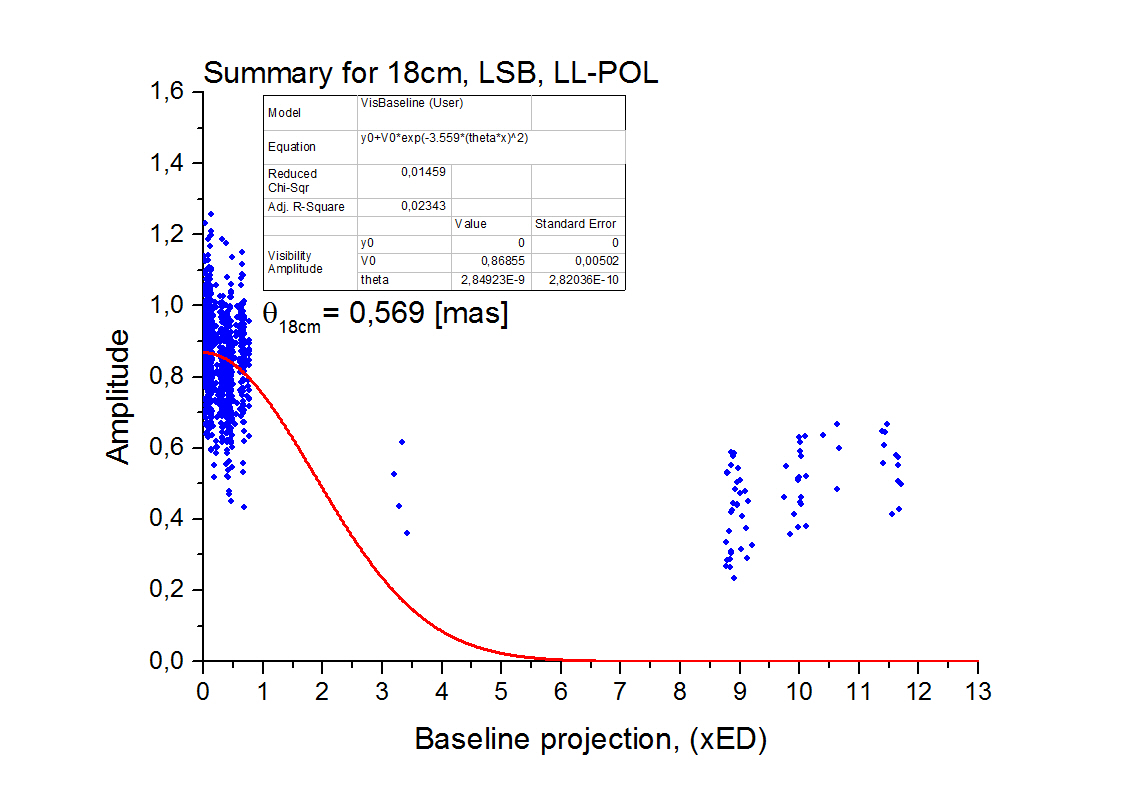}
\includegraphics[width=0.49\linewidth]{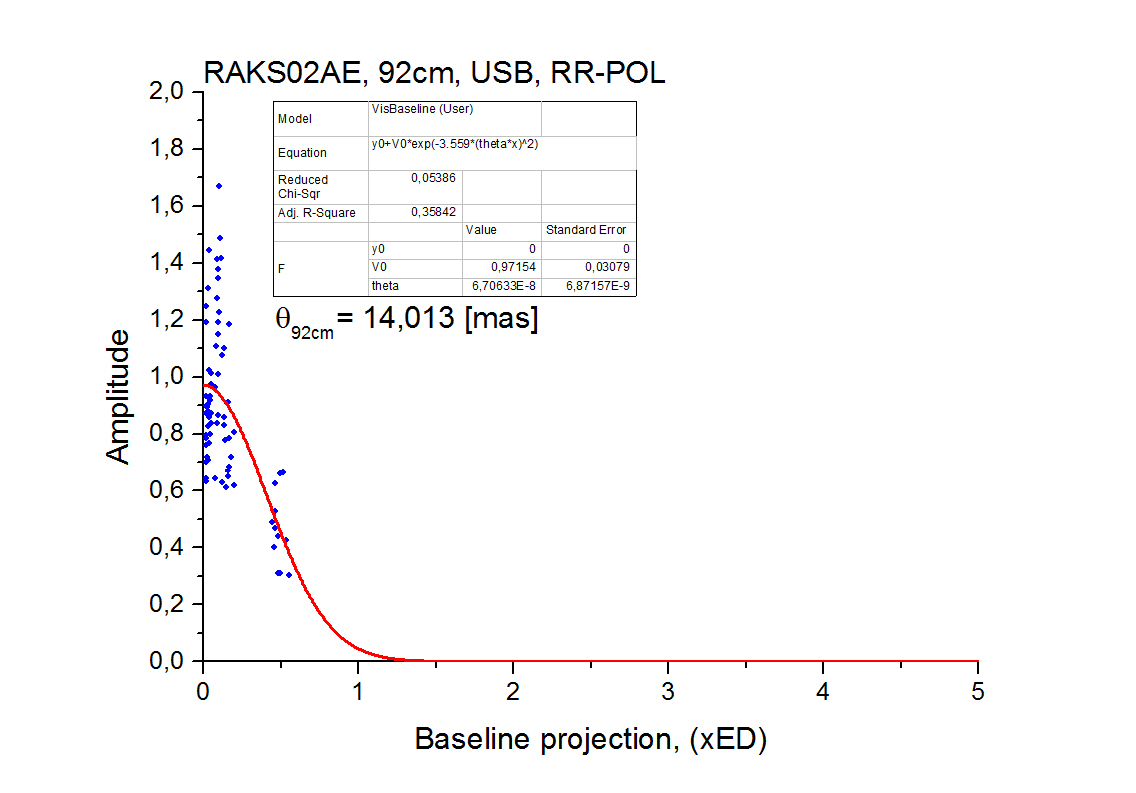}
\caption{Distribution of visibility amplitude versus baseline projection for the summary of all 18 cm observations (left) and 92 cm observation (right).}
\end{center}
\end{figure}

Distribution of visibility amplitude versus baseline projection is shown in Fig. 2. For 18 cm a summary of all observations is shown in one plot (Fig. 2, left) and for 92 cm the distribution for the only one observation done at this wavelength is shown (Fig. 2, right). At 92 cm it seems that the scattering disk is being resolved at ground baselines.

Despite our expectations, at large baseline projections ($>$ 4 Earth diameters), when the scattering disk is resolved, correlation amplitude is not equal to zero and has a significant value. What is more, the visibility amplitude stays approximately constant for maximum baseline projections ($>$ 10 Earth diameters). The same feature is observed for other pulsars \cite{ASC2}. In this article we do not discuss this effect, but we want to underline that it became definetly possible to observe this feature only with Radioastron at space-ground baselines.

\section{Summary}
All the observations within Radioastron project were successfully correlated. In four out of five sessions correlation for space-ground baselines was successfully found. We found significant change in the shape of cross-correlation function for space-ground baselines. Distribution of visibility versus baseline projection coincides with the theory of Goodman and Narayan. At 92 cm scattering disk is being resolved at ground-ground baselines. Estimated distance to the scattering screen shows the necessity to use a multiple screen model. Parameters that were estimated using the assumption of signle thin scattering screen model showed that at least two screens are required to describe the scattering situation for Crab pulsar in more details. Additionaly, 18 hours of Crab pulsar observations were approved by Radioastron AO-2 committee and by ground telescopes. We hope that the results of these observations will provide us better statistics for our studies.

\section{Acknowledgements}
The RadioAstron project is led by the Astro Space Center of the Lebedev Physical Institute of the Russian Academy of Sciences and the Lavochkin Scientific and Production Association under a contract with the Russian Federal Space Agency, in collaboration with partner organizations in Russia and other countries. Partly based on observations performed with radio telescopes of IAA RAS (Federal State Budget Scientific Organization Institue of Applied Astronomy of Russian Academy of Sciences). Partly based on observations performed with the European VLBI Network --- a joint facility of European, Chinese, South African and other radio astronomy institutes funded by their national research councils.

\end{document}